\documentclass{article}

\usepackage[svgnames]{xcolor}
\usepackage{listings}

\usepackage[utf8]{inputenc}
\usepackage{amsmath}  
\usepackage{dsfont}

\usepackage{pgfplots}
\pgfplotsset{
 compat=1.11
}

\usepackage{float}
\usepackage{multirow}
\usepackage{graphicx}
\usepackage{caption}
\usepackage{subcaption}

\usepackage{natbib}
\usepackage{amssymb}
\usepackage[colorlinks,allcolors=black,pdftex]{hyperref}

\usepackage[nottoc]{tocbibind}
\usepackage{tikz-qtree}

\usepackage[export]{adjustbox}

\usepackage{verbatim}

\title{Tree models for covariate-dependent method agreement with repeated measurements in clinical research} 

\author{Siranush Karapetyan$^{a}$, Achim Zeileis$^b$, Moritz Flick$^c$, Bernd Saugel$^c$, \\ Alexander Hapfelmeier$^{a,d,*}$} 

\date{%
$^a$ Institute of General Practice and Health Services Research, TUM School of Medicine and Health, Technical University of Munich, Munich, Germany \\%
$^b$ Faculty of Economics and Statistics, University of Innsbruck, Innsbruck, Austria \\%
$^c$ Department of Anesthesiology, Center of Anesthesiology and Intensive Care Medicine, University Medical Center Hamburg-Eppendorf, Hamburg, Germany \\%
$^d$ Institute of AI and Informatics in Medicine, TUM School of Medicine and Health, Technical University of Munich, Munich, Germany \\%
\today}

\begin{document}
\maketitle

$^*$Corresponding author. E-mail address: \href{mailto:alexander.hapfelmeier@tum.de}{alexander.hapfelmeier@tum.de}

\begin{abstract}

\vspace{0.2cm}
Background: In clinical research, the Bland-Altman analysis is commonly used to assess agreement of metric measurements made by two or more techniques, devices or methods. The approach can also deal with repeated measurements per subject or observational unit. However, a strong and implicit assumption is that agreement of methods is homogeneous across subjects or any other internal or external factors.

Objective: To extend the previously introduced multivariable modeling approach for conditional method agreement with single measurements per subject to the frequent case of repeated measurements.

Methods: Appropriate regression trees, called conditional method agreement trees (COAT), are generalized to capture the dependence of the parameters of the Bland-Altman analysis on covariates. These parameters, namely the expectation and variance of the differences between the methods are decomposed into subject-specific components to detect heterogeneity while accounting for repeated measurements. Whilst the theoretical, asymptotic properties of tree models are known, a supportive simulation study was carried out to assess the performance of COAT in finite samples. A comparison of devices measuring cardiac output serves as an application example. 

Results: The simulation study showed that COAT is applicable to the two relevant cases of paired and unpaired repeated measurements. It controlled the type-I error at the nominal level and could detect covariate-dependent method agreement with increasing sample size. The Adjusted Rand Index, a measure of concordance between the estimated and true subgroups, reached very high values of up to $0.9$, close to the maximum of $1$. The analysis of cardiac output showed that patients' characteristics and the size of measurements may have influence on the agreement between measuring devices, with implications for their use in patient care. 

Conclusion: COAT can explicitly define subgroups of heterogeneous method agreement in dependence of covariates with appropriate statistical significance testing in the case of repeated measurements. It is implemented in the free and open-source R package \texttt{coat}.

\end{abstract}


\section{Introduction}
\label{sec:intro}

In clinical medicine, it is crucial to assess whether measurements made by new technologies are equivalent or at least clinically comparable to those made by established and accurate measurement methods. For example, cardiac output (CO) is a complex hemodynamic variable that is difficult to measure because established measurement methods are invasive, can cause complications, require training, and are costly. With the development of new technologies, clinicians have the option to use alternative methods to measure CO that potentially alleviate these problems. However, before such new technologies can be implemented, their measurement performance must be assessed in method comparison studies to ensure optimal patient care and safety \citep{3_bland1999measuring, Bunce2009, hanneman2008design, saugel2025statistical}.

A systematic review of method comparison studies revealed that $85\%$ of studies used Bland-Altman (BA) analysis to evaluate method agreement \citep{zaki2012statistical}, underscoring the fact that BA analysis is an established standard \citep{ 1_altman1983measurement, 2_bland1986statistical, 3_bland1999measuring, 4_bland2007agreement}. It essentially estimates two quantities. The so-called ‘bias’, which measures the average difference of measurements made by two methods, and the 'Limits of Agreement' (LoA), which are a $95\%$ prediction interval, describing a respective range of the expected difference between the methods when used to make a single measurement of an unknown value. The LoA are derived from estimates of the bias and the standard deviation of the differences of measurements.

In our research, we question the universal and implicit assumption of BA analysis, that the agreement of methods is constant across all subjects (or any other kind of observational units), independent of internal and external factors. We therefore allow for differences of measurements to be influenced by patient characteristics, pre-existing conditions, experimental settings, and so on. For example, the agreement of CO measurements between invasive and non-invasive methods may relate to physics and other characteristics of the patients, such as age, sex, body mass index (BMI) or comorbidities. 

In an earlier work, we already introduced the concept of conditional method agreement and developed multivariable tree models for the case of single measurements per subject to test and model covariate-dependent method agreement \citep[]{karapetyan2025tree}. In the present work, we extend the concept and develop a new multivariable tree model for the very common but more complex case of repeated measurements per subject. Conditional inference trees \citep[]{10_hothorn2006unbiased} with a suitable transformation of the observed measurements are used to fulfill the purpose. Repeated measurements are accounted for by subject-specific components of the bias, the between-subject variance and the within-subject variance. The two sources of variance have already been defined by Bland and Altman, such that the new approach closely resembles the established BA analysis \citep[]{3_bland1999measuring, 4_bland2007agreement}.

The proposed approach is referred to as conditional method agreement trees (COAT), and its relevance for medical research is demonstrated by application to data from repeated CO measurements made with different methods \citep[]{morris2019using}. In addition, the ability of COAT to control the type-I error probability at a nominal level, the power to detect associations, and the ability to accurately define subgroups is investigated in simulation studies.

\section{Methods} \label{Meth}
Method agreement is characterized by two quantities, the bias and the LoA, which quantify the expected difference of the methods' measurements on average or for the majority of measurements, respectively. The LoA are a 95\% prediction interval of the differences between single measurements. The bias and LoA are obtained from a BA analysis, as we estimate the expectation $\mathbb{E}(Y)$ and the variance $\mathrm{Var}(Y)$ of a random variable $Y=A-B$. $Y$ represents the differences between measurements $A$ and $B$ made by two distinct methods. Respective estimates, which are the parameters of a BA analysis, are given by the mean and the standard deviation of the observed differences $y=a-b$ (see Section \ref{sec:BAapp} for computational details). In the context of repeated measurements, the methods are used to generate $m_{Ai}$ and $m_{Bi}$ measurements or realizations of A and B for the $i$-th subject ($i \in \{1, \ldots, n\}$). These numbers of measurements can be equal or unequal, as outlined below for the specific cases of paired and unpaired measurements. In our work, we allow the agreement of methods and therefore the parameters of the BA analysis to be dependent on covariates~$X$. Consequently, we define conditional method agreement by estimating the conditional expectation (or bias) $\mathbb{E}(Y|X)$ and conditional variance $\mathrm{Var}(Y|X)$. 

It is important to note that we use the term `subjects' to refer to an observational unit, although the more general term `item' is also commonly used. However, this purely notational difference has no impact on the approach presented here.

The following Section \ref{sec:BAapp} describes the original BA analysis for estimating the basic parameters of the BA analysis, i.e., the bias $\mathbb{E}(Y)$ and variance $\mathrm{Var}(Y)$ of the differences $Y$ in the case of repeated measurements. The novel and corresponding modeling of the conditional BA parameters $\mathbb{E}(Y|X)$ and $\mathrm{Var}(Y|X)$ is based on these calculations and is proposed in Section \ref{newapproach} with appropriate cross-references to the original approach.

\subsection{The Bland-Altman analysis} \label{sec:BAapp}
In the case of repeated measurements, Bland and Altman suggest distinguishing between two general settings: unpaired measurements, where the true value is constant and does not change between the repeated measurements, and paired measurements, where the true value varies between repeated measurements. Estimating the bias and the standard deviation of $Y$ differs depending on these settings \citep[]{3_bland1999measuring}. In the following, we focus on the estimators for bias and variance on which the present work is based, and refer to the original works by Bland and Altman for extensive and well-known details regarding their derivation \citep[]{1_altman1983measurement, 2_bland1986statistical, 3_bland1999measuring, 4_bland2007agreement}.

\subsubsection*{Unpaired measurements}
The main assumption here is that the repeated measurements of a method are interchangeable within a subject because they all measure the same invariant true value. 

In this case, the bias $\mathbb{E}(Y)$ can be estimated as 

\begin{align} \label{bias_unpaired}
    \overline{y} = \frac{1}{n}\sum_{i=1}^n \overline{y}_i,
\end{align}

where $\overline{y}_i = \overline{a}_{i} - \overline{b}_{i}$ is the difference of the mean values of measurements per method for the subject $i \in \{1, \ldots, n\}$. 

The variability $\mathrm{Var}(Y) = \sigma_Y^2$ of the differences $Y$ is estimated by

\begin{align} \label{unpair_var}
\widehat{\sigma}_Y^2=\widehat{\sigma}_{\overline{Y}}^2 + \left(1 - \frac{1}{n} \sum_{i=1}^n \frac{1}{m_{Ai}}\right)\widehat{\sigma}_{A}^2 + \left(1 - \frac{1}{n} \sum_{i=1}^n \frac{1}{m_{Bi}}\right)\widehat{\sigma}_{B}^2, 
\end{align}

where $m_{Ai}$ and $m_{Bi}$ are the numbers of measurements $A$ and $B$ for a subject $i$, respectively, and may be unequal between methods and subjects. The between-subject variance $\sigma_{\overline{Y}}^2$ is estimated as the sample variance of the mean differences per subject by

\begin{align} \label{betw_var_unpaired}
\widehat{\sigma}_{\overline{Y}}^2 = \frac{1}{n-1} \sum_{i=1}^n (\overline{y}_i - \overline{y})^2.
\end{align}

The within-subject variances $\sigma_A^2$ and $\sigma_B^2$ are random errors of the methods and are the same for all subjects. A one-way analysis of variance for the measurements of each method by subject can be used to estimate

\begin{align} \label{with_var_unpaired}
\widehat{\sigma}_A^2 = \frac{1}{n} \sum_{i=1}^n \frac{r_{Ai}^2}{df_{Ai}} && 
\text{and} &&
\widehat{\sigma}_B^2 = \frac{1}{n} \sum_{i=1}^n \frac{r_{Bi}^2}{df_{Bi}}.
\end{align}

Here, $r_{Ai}^2 = \sum_{m=1}^{m_{Ai}} (a_{im} - \overline{a}_i)^2$ and $r_{Bi}^2 = \sum_{m=1}^{m_{Bi}} (b_{im} - \overline{b}_i)^2$ are the residual sum of squares of measurements $a$ and $b$ for subject $i$. The subject-specific degrees of freedom $df_{Ai}$ and $df_{Bi}$ are $(m_{Ai}-1)$ and $(m_{Bi}-1)$, respectively.

\subsubsection*{Paired measurements}
One may also be interested in measuring the current value of a continually changing variable, such as physical activity or cardiac output. The main assumption in this case is that the repeated measurements of a method are not interchangeable within a subject because the underlying true value varies. However, the measurements are carried out in such a way that they are paired between the methods, i.e., one measurement from each of the two methods together form a measurement pair for the same unknown true value. The numbers of measurements $m_{Ai}$ and $m_{Bi}$ are therefore $m_i$ in this paired setting but may well differ between subjects. 

The bias $\mathbb{E}(Y)$ is estimated by the mean difference, weighted by the number of observations per subject

\begin{align} \label{bias_paired}
    \overline{y} = \frac{1}{\sum_{i=1}^n m_i} \sum_{i=1}^n m_i \overline{y}_i,
\end{align}

where $\overline{y}_i = \frac{1}{m_{i}}\sum_{m=1}^{m_{i}} y_{im}$ are the mean differences of paired measurements per subject. 

The variability $\mathrm{Var}(Y) = \sigma_Y^2$ of the paired differences $Y$ is estimated by

\begin{align} \label{pair_var}
    \widehat{\sigma}_Y^2 =  (\widehat{\sigma}_{\overline{Y}}^2 - \widehat{\sigma}_{w}^2) / \frac{(\sum m_i)^2 - \sum m_i^2}{(n - 1) \sum m_i} + \widehat{\sigma}_{w}^2.
\end{align}

As in \eqref{betw_var_unpaired}, the between-subject variance $\sigma_{\overline{Y}}^2$ is estimated by

\begin{align} \label{with_var_db}
\widehat{\sigma}_{\overline{Y}}^2 = \frac{1}{n-1} \sum_{i=1}^n \left(\overline{y}_i - \overline{y}\right)^2.
\end{align}

Using a one-way analysis of variance, the within-subject variance $\sigma_{w}^2$ is estimated by

\begin{align} \label{with_var_paired}
\widehat{\sigma}_{w}^2 = \frac{1}{n} \sum_{i=1}^n \frac{r_i^2}{df_i},
\end{align}

where $r_i^2 = \sum_{m=1}^{m_{i}} (y_{im} - \overline{y}_i)^2$ is the residual sum of squares for subject $i$. The subject-specific degrees of freedom $df_i$ equal $(m_i-1)$.

\subsection{Conditional method agreement trees} \label{newapproach}
We consider a tree-based algorithm, namely conditional inference trees (CTree), to estimate a bivariate outcome that includes the conditional bias $\mathbb{E}(Y|X)$ and the conditional variance $\mathrm{Var}(Y|X)$ of the differences $Y$, thereby defining heterogeneous subsets in terms of these parameters \citep[]{10_hothorn2006unbiased}. In general, a CTree uses (asymptotic) permutation test statistics to explore whether there is a statistically significant association of the outcome to the covariates and to define optimal splitting along the values or levels of the covariates into subsets with heterogeneous outcomes. More precisely, the null-hypothesis $H_0^j: f_{Y}(y|x_j) = f_{Y}(y)$ of independence of $Y$ from the $j$-th covariate $X_j$ is tested by use of the statistic 

\[
t_j = \mathrm{vec} \left( \sum_{i=1}^n \omega_i g_j(x_{ji}) h(y_i, (y_1, ..., y_n))^\top \right),
\]

which captures the association (or correlation) between a transformation $g(X_j)$ of the covariate and a transformation $h(Y)$ of the outcome. For further details, we refer to the original works on CTree and the testing framework \citep[]{strasser1999asymptotic, 10_hothorn2006unbiased}. 

Concerning COAT, it is important to emphasize the crucial role played by the non-random transformation function $h(\cdot)$. In our previous work with single measurements per method and subject, we already successfully modeled $\mathbb{E}(Y|X)$ and $\mathrm{Var}(Y|X)$ by defining the transformation function as $h(\cdot) = (y_i, (y_i - \overline{y})^2)$ \citep[]{karapetyan2025tree}. In the present work, the transformation function is adapted to each of the two cases of unpaired and paired measurements as follows below, with appropriate cross-references to the original BA analysis outlined in Section \ref{sec:BAapp}.

\subsubsection*{Unpaired measurements}
Technically, the data about the measurements made is pre-processed to compute subject-individual components of the parameters of the BA analysis. These components are handed over to the transformation function of a CTree which eventually defines the transformed outcome to be modeled. 

Analytically, all of these steps are covered by the respective definition of the transformation function $h(\cdot) = (\overline{y}_i, \widehat{\sigma}_{Yi}^2)$. It can be easily shown that this procedure is sensible to model the parameters of the BA analysis as defined in the previous Section \ref{sec:BAapp}. To comprehend this approach, it is important to bear in mind that a CTree internally calculates the average value as a summary measure of the observed outcomes contained in the data currently being evaluated for further splitting into subsets.

Concerning the bias, it is obvious that averaging the subject-individual components $\overline{y}_i$ defined in the transformation function exactly results into \eqref{bias_unpaired}. Regarding the variance, the subject-individual components of the between-subject variance are $\widehat{\sigma}_{\overline{Y}i}^2 = \frac{n}{n-1} (\overline{y}_i - \overline{y})^2$, with an average that equals \eqref{betw_var_unpaired}. The subject-individual components corresponding to the within-subject variances given in \eqref{with_var_unpaired} are $\widehat{\sigma}_{Ai}^2 = r_{Ai}^2 / df_{Ai}$ and $\widehat{\sigma}_{Bi}^2 = r_{Bi}^2 / df_{Bi}$. Consequently, 

\begin{align*}
\widehat{\sigma}_{Yi}^2 = \widehat{\sigma}_{\overline{Y}i}^2 + (1 - \frac{1}{n} \sum_{i=1}^{n} \frac{1}{m_{Ai}}) \widehat{\sigma}_{Ai}^2 + (1 - \frac{1}{n} \sum_{i=1}^{n} \frac{1}{m_{Bi}}) \widehat{\sigma}_{Bi}^2,
\end{align*}

as used in the transformation function, averages to \eqref{unpair_var}. Based on these calculations, it is possible to apply the transformation function $h(\cdot) = (\overline{y}_i, \widehat{\sigma}_{Yi}^2)$ for building the COAT model.

\subsubsection*{Paired measurements}
With paired measurements, the transformation function $h(\cdot) = (n m_i \overline{y}_i / \sum m_i, \widehat{\sigma}_{Yi}^2)$ is slightly different.

Again, the average of the subject-individual components $n m_i \overline{y}_i / \sum m_i$ of the bias clearly results to the weighted mean defined in \eqref{bias_paired}. With reference to the variance, the subject-individual components of the between-subject and within-subject variances are  $\widehat{\sigma}_{\overline{Y} i}^2 = \frac{n}{n-1} \left(\overline{y}_i - \overline{y}\right)^2$ and $\widehat{\sigma}_{w i}^2 = r_i^2 / df_i$, with averages equal to \eqref{with_var_db} and \eqref{with_var_paired}, respectively. Consequently, the average of 

\begin{align*}
\widehat{\sigma}_{Yi}^2 = (\widehat{\sigma}_{\overline{Y} i}^2 - \widehat{\sigma}_{w i}^2) / \frac{(\sum m_i)^2 - \sum m_i^2}{(n - 1) \sum m_i} + \widehat{\sigma}_{w i}^2
\end{align*}

equals \eqref{pair_var}. This again shows that $h(\cdot) = (n m_i \overline{y}_i / \sum m_i, \widehat{\sigma}_{Yi}^2)$ can be used to build the COAT model.

\subsubsection*{A two-sample test}
The proposed approach can also be used to perform a two-sample test to compare agreement between two groups. For example, in the case of the following application, one may be interested in whether agreement of methods measuring cardiac output is different between female and male patients, or between patients with or without vascular or cardiac disease. To perform such a test, it suffices to apply COAT to a corresponding binary covariate that defines these groups.

Analogously, the tests underlying COAT can be used for assessing the dependence of the BA parameters on covariates of other scales.

\section{Application}
In patients having cardiac surgery, accurate monitoring of hemodynamic variables is very important. For example, during and after surgery, arterial blood pressure is measured continuously with an arterial catheter, whereas CO can be measured using a pulmonary artery catheter \citep[]{bootsma2022contemporary1, bootsma2022contemporary2, heringlake2023classification}.
Innovative methods, such as the NiCCI system (Getinge, Gothenburg, Sweden) investigated in the present study, allow monitoring CO non-invasively without the need for inserting catheters. This system uses a finger cuff that is placed on the index and middle finger of one hand to record the arterial blood pressure waveform, which is analyzed to estimate CO.   

In the present application, we use data from a previously published study \citep{flick2022new} in which simultaneous CO measurements obtained by the NiCCI System (test method) were compared to CO measurements obtained by intermittent manual pulmonary artery thermodilution with a pulmonary artery catheter (reference method). The finger cuff measurements are hereafter referred to as CO-FC, and the pulmonary artery catheter measurements as CO-PAC. The purpose of this current study is to re-analyze the data considering the patients' characteristics and their comorbidities. For this, the agreement of CO measurements between the two technologies is compared with respect to the patients' sex, age, height, weight, BMI and comorbidities.

\subsection{Data}
A total of $n=51$ patients who had cardiac surgery at the University Medical Center Hamburg-Eppendorf, Hamburg, Germany were included in the study.  Patient characteristics are shown in Table~\ref{tab:part_char}. CO was measured in each patient using the aforementioned methods at $5$ consecutive and distinct measurement time points. At each measurement time point, up to $5$ simultaneous measurements of CO-FC and CO-PAC were performed. 

\begin{table}[ht]
    \centering
    \caption{Patient characteristics of the application study ($n=51$).}
    \begin{tabular}{ll}
    \hline
    Variables & $n$ (\%); Median (IQR) \\
    \hline
    Male & $43 \ (84\%)$ \\
    Age (years) & $72 \ (64, 78)$ \\
    Height (cm) & $175 \ (169, 180)$ \\
    Weight (kg) & $83 \ (75, 88)$ \\
    BMI & $27 \ (24, 30)$ \\
    Vascular disease & $27 \ (53\%)$ \\
    Arteriosclerosis & $45 \ (88\%)$ \\
    Diabetes & $8 \ (16\%)$ \\
    \hline
    \end{tabular}
    \label{tab:part_char}
\end{table}

\begin{figure}[t!]
    \centering
    \includegraphics[width=0.8\textwidth, trim=0 0 0 50, clip]{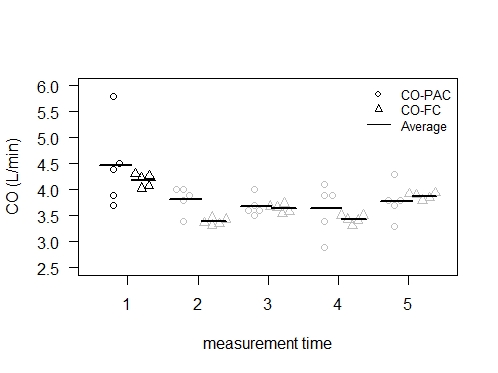}
    \caption{CO measurements for a sample patient from the dataset. At five measurement time points, five measurements were taken for each method, and mean values were calculated. Grey symbols indicate measurements that were not used further in the application. The black circles and triangles at the first measurement time point represent measurements used in the unpaired setting of the application, as they cannot be directly matched to one another. The mean values for all measurement time points, represented by horizontal lines, were used in the paired setting of the application.}
    \label{COData}
\end{figure}

As discussed in Section \ref{newapproach}, our analysis involves the two different settings of paired and unpaired measurements, both of which are supported by the data. To prepare the data for the case of paired measurements, we calculated the mean values of a patient's CO-FC and CO-PAC at each distinct measurement time point, respectively. This eventually lead to five pairs of CO-FC and CO-PAC measurements per patient for analysis. In contrast, to generate and analyze the case of unpaired measurements, we exclusively focused on the first measurement time point and compare the five simultaneous CO-FC and CO-PAC measurements taken at that point. For better clarity, a graphical representation of the measurements is provided in Figure~\ref{COData}.

\subsection{Implementation}
We employ COAT as proposed in Section~\ref{newapproach} and as implemented in the R package \texttt{coat} version 0.3.0, which is provided as supplementary material. The values \texttt{minsize} $=6$ and \texttt{alpha} $=0.05$ (default) of the algorithm are used to determine the minimal size of a defined subset and the significance level for splitting. We include the aforementioned characteristics of the patients as covariates, with and without use of mean CO measurements as additional covariate. In some cases, these settings are relaxed to allow for more splits and a deeper exploratory investigation of the potential dependence of agreement on the covariates. Such modifications are indicated along the presentation of respective results.

\subsection{Results}

\subsubsection*{Unpaired measurements}
\begin{figure}[t!]
    \centering
    \includegraphics[width=0.85\textwidth, trim=0 0 0 15, clip]{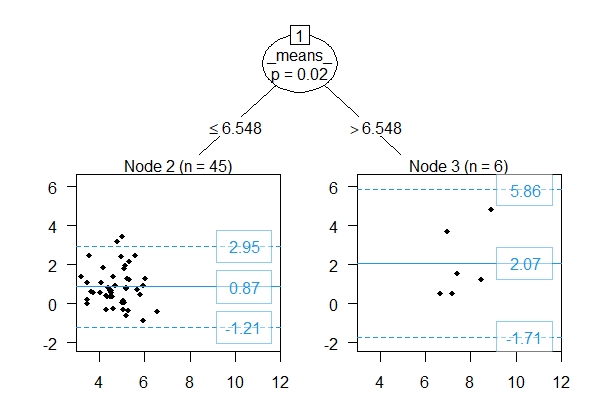}
    \caption{COAT for conditional agreement of cardiac output (CO) measurements of two different technologies. CO-PAC and CO-FC are compared for the unpaired measurements of CO. The minimal subgroup size (\texttt{minsize} $= 6$) and the significance level ($\alpha = 0.05$) are set as model parameters. The mean measurements are included along with other patient characteristics as covariates.}
    \label{COATunpaired}
\end{figure}

In order to explore unpaired CO measurements, we take the first measurement time point and compare the multiple measurements carried out at this time. Applying COAT to compare CO-PAC and CO-FC reveals a statistically significant difference in agreement with regard to mean measurements (Figure~\ref{COATunpaired}), while patient characteristics were not chosen for splitting. In particular, better agreement in terms of bias and width of the LoA is achieved with lower measurement values $\leq 6.548$ L/min.

\begin{figure}[t!]
    \centering
    \includegraphics[width=\textwidth]{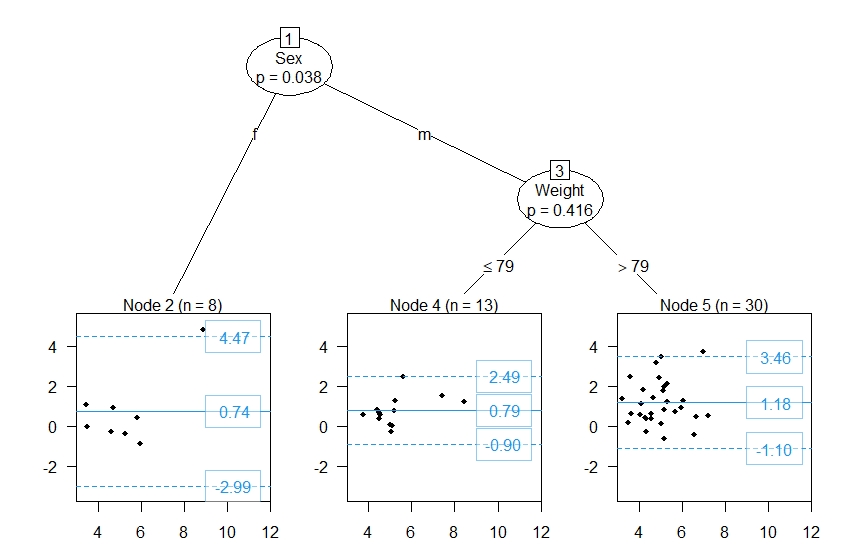}
    \caption{COAT for conditional agreement of cardiac output (CO) measurements of two different technologies. CO-PAC and CO-FC are compared for the unpaired measurements of CO. The minimal subgroup size (\texttt{minsize} $= 6$) and the significance level ($\alpha = 1$) are set as model parameters. The tree growth was restricted to a maximal depth of two (\texttt{maxdepth} $= 2$). The mean measurements are not included as covariate.}
    \label{COATunpaired_depth}
\end{figure}

We repeated the analysis without using mean measurements as a covariate and setting $\alpha =1$ and \texttt{maxdepth} $= 2$, in order to enable an exploration of the first potential associations to patient characteristics. COAT detects a statistically significant difference in agreement between female and male patients (Figure \ref{COATunpaired_depth}). In male patients, the best agreement is observed in those weighing 79 kg or less, with regard to the bias and the width of the LoA. However, this exploratory result lacks statistical significance, and further insights into this finding are required.

\subsubsection*{Paired measurements}
\begin{figure}[p!]
    \centering
    \includegraphics[width=0.8\textwidth, trim=0 0 0 15, clip]{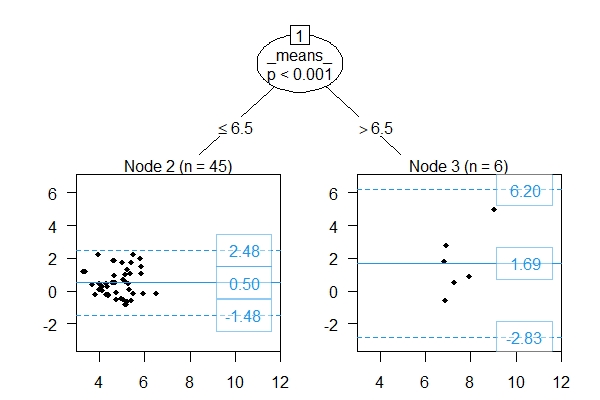}
    \caption{COAT for conditional agreement of cardiac output (CO) measurements of two different technologies. CO-PAC and CO-FC are compared for the paired measurements of CO. The minimal subgroup size (\texttt{minsize} $= 6$) and the significance level ($\alpha = 0.05$) are set as model parameters. The mean measurements are included along with other patient characteristics as covariates.}
    \label{COATpaired}

    \includegraphics[width=\textwidth, trim=0 0 0 5, clip]{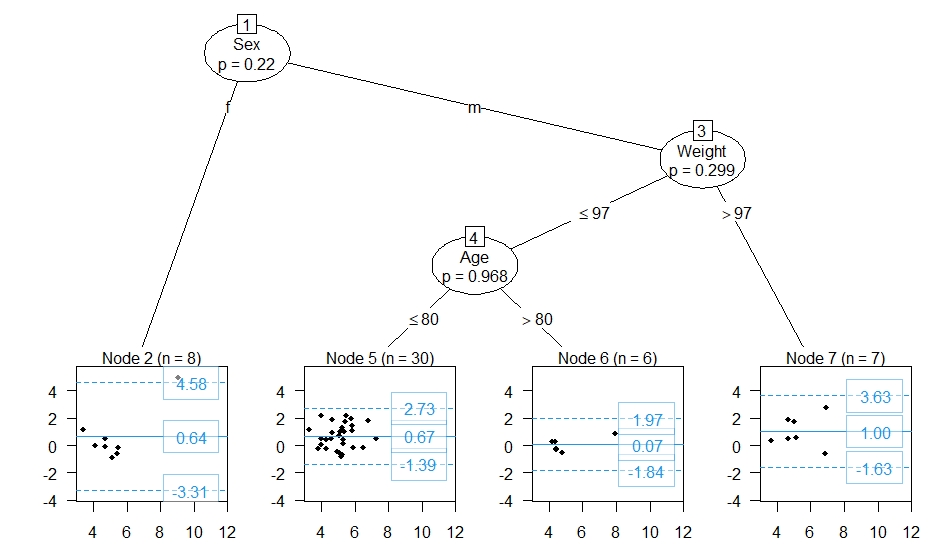}
    \caption{COAT for conditional agreement of cardiac output (CO) measurements of two different technologies. CO-PAC and CO-FC are compared for the paired measurements of CO. The minimal subgroup size \texttt{minsize} $= 6$ and the significance level $\alpha = 1$ are used as model parameters to explore the dependence of the agreement on the covariates in depth. The tree growth was restricted by choosing \texttt{maxdepth} $= 3$. The mean measurements are not included as covariate.}
    \label{COATpaired_depth}
\end{figure}

Mean values per measurement time and technology are used in this analysis to explore the case of paired CO measurements. Similar to the previous application to unpaired measurements, it can be seen that better agreement in terms of both BA parameters is achieved with lower mean measurements $\leq 6.500$ L/min (Figure~\ref{COATpaired}). 

Focusing on patient characteristics only, i.e., excluding mean values as a covariate, the variables Sex, Weight and Age also play a certain role (Figure~\ref{COATpaired_depth}). This exploratory result was obtained using $\alpha=1$ and restricting the tree growth to \texttt{maxdepth} $=3$ subsequent splits. It is not statistically significant and should be interpreted with caution.

\subsubsection*{Two-sample BA test}

\begin{figure}[t!]
    \centering
    \includegraphics[width=0.8\textwidth, trim=0 0 0 30, clip]{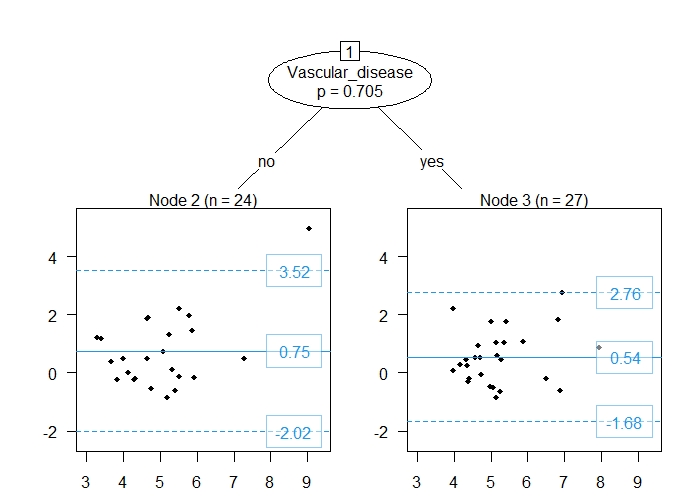}
    \caption{Two-sample BA test of differences in method agreement of cardiac output (CO) measurements between patients with and without vascular disease. CO-PAC and CO-FC are compared. Setting $\alpha = 1$ ensures that the test result is provided independent of the size of the respective p-value.}
    \label{COATpaired_vascular}
\end{figure}

As mentioned in Section~\ref{newapproach}, COAT can also be used to perform a two-sample test to compare the agreement between (pre)defined subgroups. In this application, one may be interested in the potential difference in agreement between patients with or without vascular disease. Therefore, COAT was also applied to vascular disease as the only binary covariate using the example of paired measurements. To ensure that the test result is displayed regardless of it's significance, $\alpha = 1$ was set in the implementation. The test on vascular disease produces a p-value of $p = 0.705$ (Figure~\ref{COATpaired_vascular}). 

\section{Simulation study}

\subsection{Design}
We conducted a simulation study to investigate the performance of COAT in terms of type-I error and the power to detect existing associations to covariates and respective subgroups. As a benchmark, we also include a CTree with the default transformation function to model the average difference between methods, i.e. $h(\cdot)=\overline{y}_{i}$. The minimal subgroup size, i.e. \texttt{minsize} in COAT and \texttt{minbucket} in CTree, and the significance level \texttt{alpha}, correspond to the default values of $10$ and $0.05$. The minimal subgroup size considered for further splitting was set to \texttt{minsplit} $= 20$, which is also the default in CTree. These choices apply to all models for a fair comparison.

Choices on the number of subjects ($n$) and replicates ($m_{i}$) in the simulated data is guided by real-world method comparison studies \citep[]{henriksen2019validity, flick2022new}. Common sample sizes in these types of studies are considered using $n \in \{50, 75, 100, \ldots, 300\}$ and $m_{i} = 3$ replicates for each subject. Three scenarios are investigated as follows, each comprising $5000$ simulations. 

In the ``Null'' scenario, five normally distributed covariates $\mathbf{X} \sim \mathcal{N}(\mathbf{\mu}, \mathbf{\sigma})$, with $\mathbf{\mu} = (20, 100, 5000, 0, 1000)^\top$ and $\mathbf{\sigma} = diag(4, 20, 100, 1, 50)$ are chosen to reflect the distribution of covariates in the aforementioned real case studies. These uncorrelated covariates are defined as non-informative, i.e., they have no relation to the bias and the variance of the differences $Y$. In the simulation, the bias and variance have been chosen to reflect values found in a real case with systolic blood pressure measurements \citep[]{altman1991analysis}. In particular, they are $\mathbb{E}(Y)=16$ and $\mathrm{Var}(Y) = 439$ with the variance components $\sigma_{\overline{Y}}^2 = 361$, $\sigma_{A}^2 = 36$ and $\sigma_{B}^2 = 81$ for the case of unpaired measurements. Corresponding data was generated using the function \texttt{Meth.sim()} in the \texttt{R} package \texttt{MethComp} \citep[]{carstensen2020package}, where repeated measurements are produced through a mixed-effects model as proposed by \cite{5_carstensen2004comparing}. For the case of paired measurements, the mixed-effects model additionally contains a random effect reflecting the variance of replicates within subjects. This quantity is needed to properly simulate the data, but cancels out in the computation of method agreement. Therefore, we simply adopted the default value of $6.25$ of the function \texttt{Meth.sim()} \citep[]{7_carstensen2008statistical}. The ``Null'' scenario allows us to examine the type-I error as we count statistically significant p-values in the root nodes of the COAT models fitted to the simulated data. It is important to note here that CTree and COAT use Bonferroni correction for multiple testing by default. The nominal significance level is therefore set to $\alpha=0.05$.

The ``Stump'' scenario covers two different cases, each with both unpaired and paired measurements considered as explained for the null scenario. In each of the cases, there are again five normally distributed and uncorrelated covariates $\mathbf{X} \sim \mathcal{N}(\mathbf{\mu}, \mathbf{\sigma})$ as defined before, but with the agreement depending on $X_{1}$. The difference between the two considered cases is determined by the dependence of the individual BA parameters on the covariate. In the first scenario, the covariate only influences the bias, while the variance of the differences remains unaffected. Consequently, the subgroups only differ in $\mathbb{E}(Y|X)$. In the second scenario, the influence of the covariate is limited to a change in the variance $\mathrm{Var}(Y|X)$, while the bias remains unaffected. As a result, only the width of the LoA differs in the subgroups. The corresponding definition is

$$
\left( \mathbb{E}(Y|X), \mathrm{Var}(Y|X) \right) = \begin{cases}
			(5 + 2 \cdot I(X_{1} \leq Q_{1,0.5}), 4) & \text{in case 1,}\\
            (5, 4 + 2 \cdot I(X_{1} \leq Q_{1,0.5})) & \text{in case 2,}
		 \end{cases}
$$

where $Q_{1,0.5} = 20$ is the $50$th percentile of the distribution of $X_1$ and is selected as a split point in $X_1$ to implement two subgroups with different agreement.

Finally, in the more complex ``Tree'' scenario, we consider two informative covariates, $X_{1}$ and $X_{2}$, and three uninformative covariates. These variables again follow the normal distribution $\mathbf{X} \sim \mathcal{N}(\mathbf{\mu}, \mathbf{\sigma})$ outlined above. The two informative variables influence the bias, variance or both parameters in three partly nested subgroups according to 

$$ \left( \mathbb{E}(Y|X), \mathrm{Var}(Y|X) \right) = (5 + 2 \cdot I(X_{1} \leq Q_{1,0.5}) \cdot I(X_{2} \geq Q_{2,0.5}), 4 + 2 \cdot I(X_{1} > Q_{1,0.5}) ).$$ 

Here, $Q_{1,0.5} = 20$ and $Q_{2,0.5} = 100$ are the $50$th percentiles of the distributions of $X_1$ and $X_2$, respectively. Both cases of unpaired and paired measurements are again considered.

The performance of COAT in the ``Stump'' scenario is assessed in terms of its power to reject the null-hypothesis of independence for the informative covariate in the root nodes of the tree models. It is estimated by the respective relative frequencies of statistically significant test results for the covariate. However, estimates of power do not indicate whether the given subgroups are correctly specified by the models. Therefore, another performance measure is applied in the ``Stump'' scenario and the ``Tree'' scenario. The Adjusted Rand Index (ARI) measures the similarity between two classifications \citep[]{hubert1985comparing}, as it quantifies the proportion of paired observations belonging to the same or different subgroups in both classifications out of the total number of paired observations \citep[]{rand1971objective}. In the case of independent or random classifications, the ARI takes a value of $0$. Higher values indicate a higher similarity, with $1$ indicating that the classifications coincide. Here, the ARI is used to evaluate the similarity between the true subgroups and the subgroups defined by COAT.

\subsection{Results}

\begin{figure}[t!]
    \centering
    \begin{subfigure}[b]{0.49\textwidth}
    \includegraphics[width=\textwidth,valign=c]{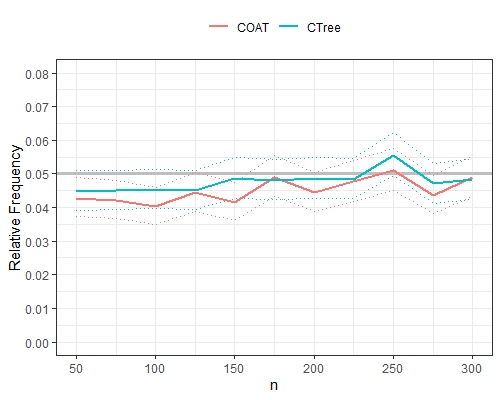}
    \subcaption[]{Unpaired measurements}
    \label{fig:sim_null1}
    \end{subfigure}
    \begin{subfigure}[b]{0.49\textwidth}
    \includegraphics[width=\textwidth,valign=c]{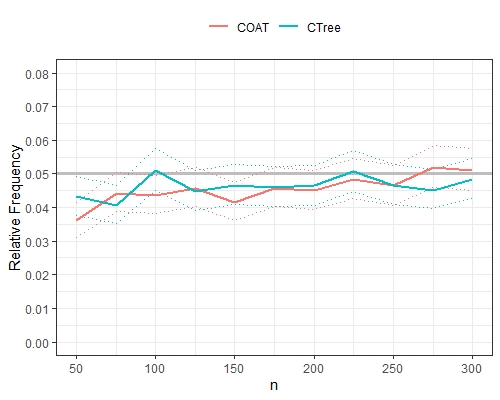}
    \subcaption[]{Paired measurements}
    \label{fig:sim_null2}
    \end{subfigure}
    \caption{Relative frequency of statistically significant p-values observed in the root nodes of COAT and CTree fit to data of increasing sample size in the ``Null'' scenario with $5000$ replications. These estimates of the type-I error probability are presented with pointwise $95\%$ confidence intervals (dashed lines).}
    \label{fig:sim_null}
\end{figure}

In the ``Null'' scenario of no association between the covariates and method agreement with unpaired measurements, the relative rejection frequencies of the null hypothesis of independence range from $4.5\%$ to $5.6\%$ for CTree, and from $4.0\%$ to $5.1\%$ for COAT with different sample sizes (Figure~\ref{fig:sim_null}). With paired measurements, the relative rejection frequencies range from $4.1\%$ to $5.1\%$ for CTree, and from $3.6\%$ to $5.2\%$ for COAT. These results indicate that both models are able to control the type-I error on the nominal significance level~of~$0.05$

In the following, the results of paired and unpaired measurements are presented, but are not discussed separately, due to their broad similarity.

\begin{figure}[t!]
    \centering
    \begin{subfigure}[b]{0.45\textwidth}
    \includegraphics[width=1\textwidth,valign=c]{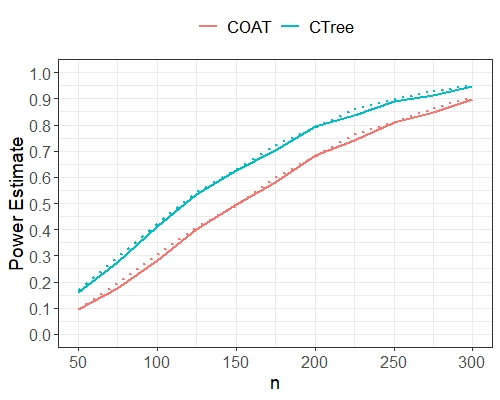}
    \subcaption[]{Power est., diff. in bias (case 1)}
    \label{fig:sel1}
    \end{subfigure}
    \begin{subfigure}[b]{0.45\textwidth}
    \includegraphics[width=1\textwidth,valign=c]{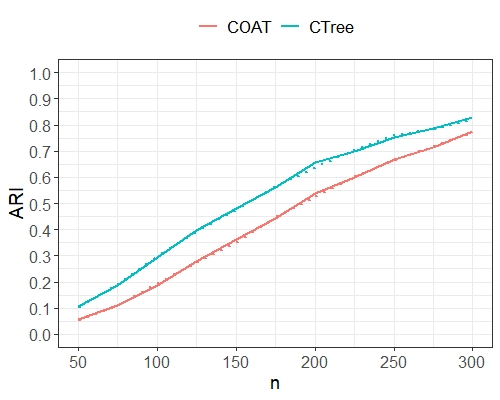}
    \subcaption[]{ARI, diff. in bias (case 1)}
    \label{fig:ARI1}
    \end{subfigure}
    \begin{subfigure}[b]{0.45\textwidth}
    \includegraphics[width=1\textwidth,valign=c]{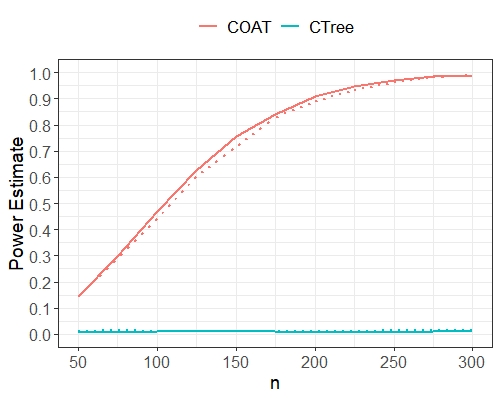}
    \subcaption[]{Power est., diff. in LoA (case 2)}
    \label{fig:sel2}
    \end{subfigure}
    \begin{subfigure}[b]{0.45\textwidth}
    \includegraphics[width=1\textwidth,valign=c]{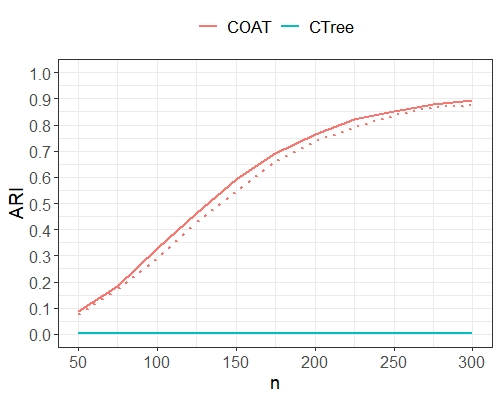}
    \subcaption[]{ARI, diff. in LoA (case 2)}
    \label{fig:ARI2}
    \end{subfigure}
    \caption{Power estimates and Adjusted Rand Index (ARI) in the ``Stump'' scenarios for increasing sample size. The ARI measures the concordance of the subgroups detected by COAT and the true underlying subgroups on a range from 0 (= random concordance) to 1 (= perfect concordance). These estimates are presented for both cases of unpaired (solid lines) and paired measurements (dashed lines).}
    \label{Eval2}
\end{figure}

In case 1 of the ``Stump'' scenario, there is only an association between the informative covariate and method agreement with regard to the bias. Here, CTree performed better than COAT in terms of the estimated power and ARI (Figure~\ref{Eval2}). In this case, a standard CTree is perfectly adequate for investigating the relationship of the covariate and the average difference between methods. COAT is slightly inferior because it examines the two parameters of the BA analysis simultaneously, whereas CTree examines only a single parameter, which coincides with the only relevant parameter in the considered case 1. Consequently, the $\chi^2$ test performed by COAT to determine statistical significance concerns a bivariate outcome and uses twice as many degrees of freedom as the test performed by CTree \citep[here it is 2 d.f.\ vs.\ 1 d.f.; cf.][]{10_hothorn2006unbiased}, which explains the lower power of COAT in the present case 1. By contrast, in case 2 of the ``Stump'' scenario, CTree completely fails to detect the association of the covariate with the LoA and is greatly outperformed by COAT. In all cases, the performance measures tend to increase with increasing sample size, with CTree being an exception when it comes to detecting associations with the LoA.

\begin{figure}[t!]
    \centering
    \includegraphics[width=0.6\textwidth,valign=c]{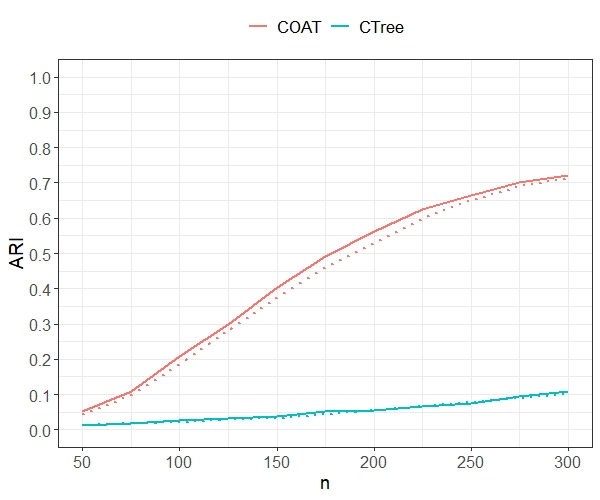}
    \label{fig:sim_tree1}
    \caption{Adjusted Rand Index (ARI) of COAT and CTree in the ``Tree'' scenario for increasing sample size. The ARI measures the concordance of the subgroups detected by COAT and the true underlying subgroups on a range from $0$ (= random concordance) to $1$ (= perfect concordance). These estimates are presented for both cases of unpaired (solid lines) and paired measurements (dashed lines).}
    \label{fig:sim_tree}
\end{figure}

In the more complex ``Tree'' scenario, there is a nested association of two informative covariates with the bias and the LoA. Here, CTree is clearly outperformed by COAT as the sample size increases (Figure~\ref{fig:sim_tree}). The results of the ``Stump'' scenario suggest that this is due to the ability of COAT to model both parameters of the BA analysis, whereas CTree can only partially capture the relation to the bias. 

\begin{figure}[ht]
    \centering
    \includegraphics[width=\textwidth,valign=c]{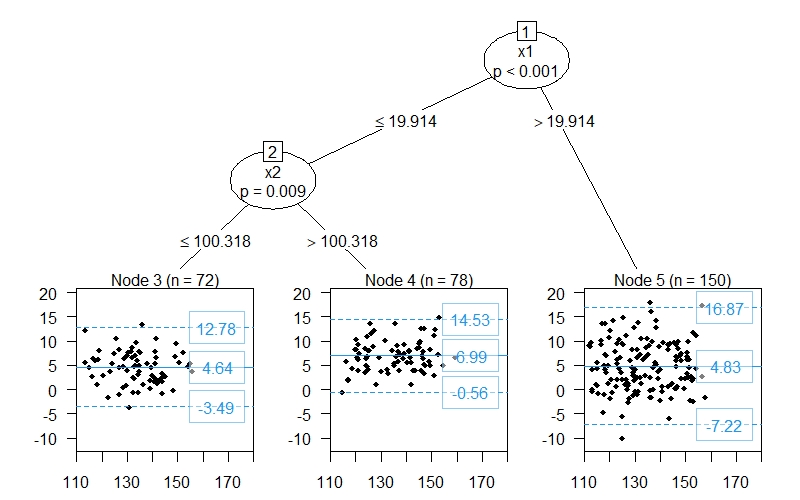}
    \caption{COAT tree fit to simulated data of the ``Tree'' scenario with paired measurements and $n=300$ subjects.}
    \label{fig:tree_example}
\end{figure}

An example of a COAT tree fit to the data of the ``Tree'' scenario with paired measurements and $n=300$ subjects can be found in Figure~\ref{fig:tree_example}. It correctly splits in $X_2$ after splitting in $X_1$ to represent the simulated association of $X_2$ to agreement conditional on $X_1$. It also closely identified the true split values $20$ and $100$ in both covariates $X_1$ and $X_2$. The estimated bias and LoA are close to the values used in the data generation. Per definition, $(\mu_{Y|X}, \sigma_{Y|X}) = (5, 4)$ in Node 3, $(7, 4)$ in Node 4 and $(5, 6)$ in Node 5, for which COAT estimated $(4.64, 4.15)$, $(6.99, 3.85)$ and $(4.83, 6.15)$.

\section{Discussion}
In the present work, we propose modeling of covariate-dependent method agreement by integrating the parameters of the BA analysis for repeated measurements into the framework of recursive partitioning. Conditional inference trees with appropriate transformations of the repeated measurements are used for this purpose. They essentially explore relations of subject-specific components of the parameters of the BA analysis, i.e. the bias and variance of differences between methods, to the observed values of the covariates. The novel modeling approach is referred to as conditional method agreement trees (COAT) and is made available through the supplementary material and the free and open-source R package \texttt{coat} \citep[]{hapfelmeier_coat_package}. The case of single measurements per subject has already been covered by our previous work \citep[]{karapetyan2025tree}.

The theoretical foundation and rationality of COAT is supported by it's analogy to the original BA analysis and by use of conditional inference trees, whose theoretical properties are well known and apply to COAT \citep{10_hothorn2006unbiased}. Despite these known properties, a simulation study was carried out which showed that COAT is able to control the type-I error at the nominal level. In addition, it also performed well in identifying the given subgroups, in particular for larger sample sizes. Our results also show that COAT tends to perform better than CTree in cases where the LoA are related to the covariates, while the opposite is true when only the bias varies between subgroups. This can be attributed to the fact that the hypothesis tests of COAT are based on two parameters, which involves twice as many degrees of freedom, compared to one parameter in CTree. This leads to a lower statistical power of COAT in the particular case of heterogeneous biases only. However, testing relations between covariates and LoA is essential and cannot be omitted, as the LoA are the most meaningful result of a BA analysis.

There are also some limitations of the present work to be mentioned. Further studies with more extensive simulations are mandated to explore the performance of COAT in dependence of differing sample sizes, varying numbers of measurements per subject, different settings of heterogeneous method agreement in terms of bias and LoA, and more complex or simply other relations to multiple covariates and respective subgroups. However, such investigations were beyond the scope of the present work, which introduces COAT for repeated measurements in theory and practice. COAT also covers both cases of paired and unpaired measurements \citep{7_carstensen2008statistical, 4_bland2007agreement}. While it is theoretically possible, and actually supported by the data of our application example, the practically less relevant case of combining paired and unpaired measurements is not dealt with methodologically \citep{hapfelmeier2016cardiac}. Future developments could further address this case.

The application examined in the present work shows that the agreement of cardiac output measurements between the invasive and non-invasive technologies may depend on the characteristics of patients. In particular, sex, age, weight and the size of measurements could alter the agreement between the techniques. This important finding shows the potential of COAT in clinical research, suggesting that specific techniques may or may not be applicable to obtain meaningful measurements in certain patients or measurement settings. It was also shown, that COAT can be applied to perform a two-sample BA test on the difference of agreement between defined groups.

\subsection*{Conclusion}
COAT for repeated measurements is a theoretically well-founded multivariable approach to model covariate-dependent method agreement. It explicitly defines subgroups of heterogeneous agreement in terms of bias and LoA and provides a test for statistical significance. As a special applicaton, COAT can also be used to perform a two-sample test for differences in agreement between defined groups.

\section*{Data availability and R Code}
The data underlying the application study can be requested from the authors of the original study upon reasonable request \citep[]{morris2019using}. The code for the application and simulation studies is included as supplementary material. COAT is also accessible through the supplementary material and the associated R package \texttt{coat}, available on the Comprehensive R Archive Network (CRAN) \citep[]{hapfelmeier_coat_package}.

\section*{Conflicts of interest}
MF is a consultant for Edwards Lifesciences (Irvine, CA, USA) and Vygon (Écouen, France) and has received honoraria for consulting and giving lectures from CNSystems Medizintechnik (Graz, Austria). BS is a consultant for Edwards Lifesciences (Irvine, CA, USA), Philips North America (Cambridge, MA, USA), GE Healthcare (Chicago, IL, USA), Vygon (Aachen, Germany), Masimo (Neuchâtel, Switzerland), Retia Medical (Valhalla, NY, USA), Maquet Critical Care (Solna, Sweden), Pulsion Medical Systems (Feldkirchen, Germany), Dynocardia (Cambridge, MA, USA), RDS (Strasbourg, France). BS has received institutional restricted research grants from Edwards Lifesciences, Baxter (Deerfield, IL, USA), GE Healthcare, Masimo, Philips Medizin Systeme Böblingen (Böblingen, Germany), CNSystems Medizintechnik (Graz, Austria), Pulsion Medical Systems, Vygon, Retia Medical, Osypka Medical (Berlin, Germany). BS has received honoraria for giving lectures from Edwards Lifesciences, Philips Medizin Systeme Böblingen, Baxter, GE Healthcare, Masimo, CNSystems Medizintechnik, Getinge (Gothenburg, Sweden), Pulsion Medical Systems, Vygon, Ratiopharm (Ulm, Germany). BS is an Editor of the British Journal of Anaesthesia. The remaining authors have no conflicts of interest to declare.

\section*{Funding}
This study was supported by the Deutsche Forschnugsgemeinschaft (DFG, German Research Foundation) [grant number $447467169$].

\section*{Author contributions}
S.K. and A.H. developed the modeling method, drafted the manuscript, performed the statistical analyses and interpreted the results. MF and BS extracted and prepared the data used in the application study. All authors contributed to the development of methods and their application, interpreted the results, revised the manuscript for its content and approved the submission of the final manuscript.

\newpage
\bibliographystyle{apalike2}
\bibliography{myref}

\end{document}